\documentclass[12pt,preprint]{aastex62} 
\usepackage{color}
\usepackage{hhline}
\usepackage{graphicx}
\usepackage{booktabs}
\usepackage{threeparttable}
\usepackage{longtable}
\usepackage{hyperref}
\usepackage{scalefnt}
\hypersetup{
    colorlinks=true,
    linkcolor=blue,
    filecolor=magenta,      
    urlcolor=blue,
}
\shorttitle{Fluorine Abundances in the Globular Cluster M 4}
\shortauthors{Guer\c{c}o et al.}

\begin{document}

\title{Fluorine Abundances in the Globular Cluster M 4}

\author{Rafael Guer\c{c}o}
\affil{Observat\'orio Nacional, S\~ao Crist\'ov\~ao, Rio de Janeiro, Brazil}
\email{rguerco@on.br}

\author{Katia Cunha}
\affil{University of Arizona, Tucson, AZ 85719, USA}
\affil{Observat\'orio Nacional, S\~ao Crist\'ov\~ao, Rio de Janeiro, Brazil}

\author{Verne V. Smith}
\affil{National Optical Astronomy Observatories, Tucson, AZ 85719, USA}

\author{Claudio B. Pereira}
\affil{Observat\'orio Nacional, S\~ao Crist\'ov\~ao, Rio de Janeiro, Brazil}

\author{Carlos Abia}
\affil{Depto. Fisica Teorica y del Cosmos, Universidad de Granada, E-18071 Granada, Spain}

\author{David L. Lambert}
\affil{W. J. McDonald Observatory and Department of Astronomy, University of Texas at Austin, Austin, TX 78712, USA}

\author{Patrick de Laverny}
\affil{Université C\^ote d'Azur, Observatoire de la C\^ote d'Azur, CNRS, Laboratoire Lagrange, France}

\author{Alejandra Recio-Blanco}
\affil{Université C\^ote d'Azur, Observatoire de la C\^ote d'Azur, CNRS, Laboratoire Lagrange, France}

\author{Henrik J\"onsson}
\affil{Lund Observatory, Department of Astronomy and Theoretical Physics, Lund University, Box 43, SE-22100 Lund, Sweden}

\begin{abstract}
 
We present chemical abundances for the elements carbon, sodium, and fluorine in 15 red giants of the globular cluster M 4, as well as six red giants of the globular cluster $\omega$ Centauri. The chemical abundances were calculated in LTE via spectral synthesis.
The spectra analyzed are high-resolution spectra obtained in the near-infrared region around $\lambda$2.3$\mu$m with the Phoenix spectrograph on the 8.1m Gemini South Telescope, the IGRINS spectrograph on the McDonald Observatory 2.7m Telescope, and the CRIRES spectrograph on the ESO 8.2m Very Large Telescope.
The results indicate a significant reduction in the fluorine abundances when compared to previous values from the literature for M 4 and $\omega$ Centauri, due to a downward revision in the excitation potentials of the HF(1-0) R9 line used in the analysis. 
The fluorine abundances obtained for the M 4 red giants are found to be anti-correlated with those of Na, following the typical pattern of abundance variations seen in globular clusters between distinct stellar populations.  In M 4, as the Na abundance increases by $\sim$+0.4 dex, the F abundance decreases by $\sim$-0.2 dex.  A comparison with abundance predictions from two sets of stellar evolution models finds that the models predict somewhat less F depletion ($\sim$-0.1 dex) for the same increase of +0.4 dex in Na.

\end{abstract}


\section{Introduction}

After H and $^{4}$He, the most abundant elements in the Galaxy are the stable nuclei of O (Z=8), Ne (Z=10), C (Z=6), and N (Z=7).  Nestled within this group, fluorine (Z=9), is some 10$^{4}$ to 10$^{5}$ times less abundant, consisting of one stable isotope, $^{19}$F.
Nonetheless, it is an astrophysically interesting element, as fluorine is one member of a group of elements that participates in the various cyclic H-burning reactions (the CN, ON, NeNa, and MgAl cycles) that have been observed to have altered the chemical compositions of the stellar populations within globular clusters (Gratton et al. 2012).  Fluorine 
falls in-between the stable nuclei involved in the ON-cycle ($^{14}$N, $^{15}$N, $^{16}$O, $^{17}$O, $^{18}$O) and the NeNa-cycle ($^{20}$Ne, $^{21}$Ne, $^{22}$Ne, $^{23}$Na).  In addition to the many studies of globular cluster abundance variations in C, N, O, Na, Mg, and Al, fluorine provides additional information on the stellar nucleosynthesis that has driven these abundance variations. 

This study focuses on the behavior of $^{19}$F (hereafter F) in the globular clusters M 4, as well as a few stars in $\omega$ Centauri. In addition, the abundances of the important elements carbon and sodium are also derived, as these also are altered by the H-burning cycles.
There have been relatively few fluorine abundance studies in globular clusters; these include $\omega$ Centauri (Cunha et al. 2003), M 4 (Smith et al. 2005; de Laverny \& Recio-Blanco 2013a), NGC 6712 (Yong et al. 2008), 47 Tuc (de Laverny \& Recio-Blanco 2013a), and M22 (Alves-Brito et al. 2012; D'Orazi et al. 2013; de Laverny \& Recio-Blanco 2013b).  
In the globular clusters M 4 and NGC 6712, F abundances were found to exhibit significant abundance variations that correlate with O abundances and anti-correlate with Na abundances; these variations would indicate that F abundances in globular clusters have been altered by the H-burning cycles discussed above. On the other hand, for 47 Tuc, an anti-correlation between Na and F has not been verified and for the globular cluster M 22 there have been conflicting results, given the difficulty in measuring weak HF lines in regions of telluric contamination. Nonetheless, fluorine provides an additional constraint on the masses of the stars that have driven the elemental abundance patterns observed in globular clusters and its abundance is thus worth pursuing in these environments.

A current issue with some published fluorine abundances is that many of the studies (particularly the earlier ones) used inconsistent energies for the lower excitation potentials of HF lines when compared to the dissociation energy. Jorissen et al. (1992; originally from Tipping 1988, private communication) used the minimum of the HF potential energy curve as the zero-energy reference level for the vibrational-rotational energy levels ($\chi$), but used a dissociation energy, D$_{0}$, that was defined relative to the lowest bound energy level.  The difference amounts to 0.253eV and was first noted in Lucatello et al. (2011) and discussed in detail by J\"onsson et al. (2014a). Although this change in $\chi$ is unlikely to remove the observed fluorine abundance correlations / anti-correlations with O and Na detected in globular clusters, it will change the absolute F abundances and resulting values of [F/O] or [F/Na].  The purpose of this study is to reanalyze and update F abundances in the globular clusters M 4 and $\omega$ Centauri using the correct excitation potential for the HF  (1-0) R9 line in both previously analyzed spectra from Cunha et al. (2003), Smith et al. (2005), and de Laverny \& Recio-Blanco (2013a), as well as new spectra obtained for additional M 4 and $\omega$ Centauri red giants.

This paper is organized as follows: Section 2 describes the observations, Sections 3 and 4 discuss the determination of the stellar parameters and the abundance analysis, respectively, Section 5 presents the results and compares the derived abundances with other results from the literature and Section 6 discusses the fluorine, carbon and sodium abundance results in terms of signatures of H-burning cycles. Finally, the conclusions are found in Section 7.

\section{Observations}
    
\begin{figure}[!htb]
   \centering
   \includegraphics[width=1.0\textwidth]{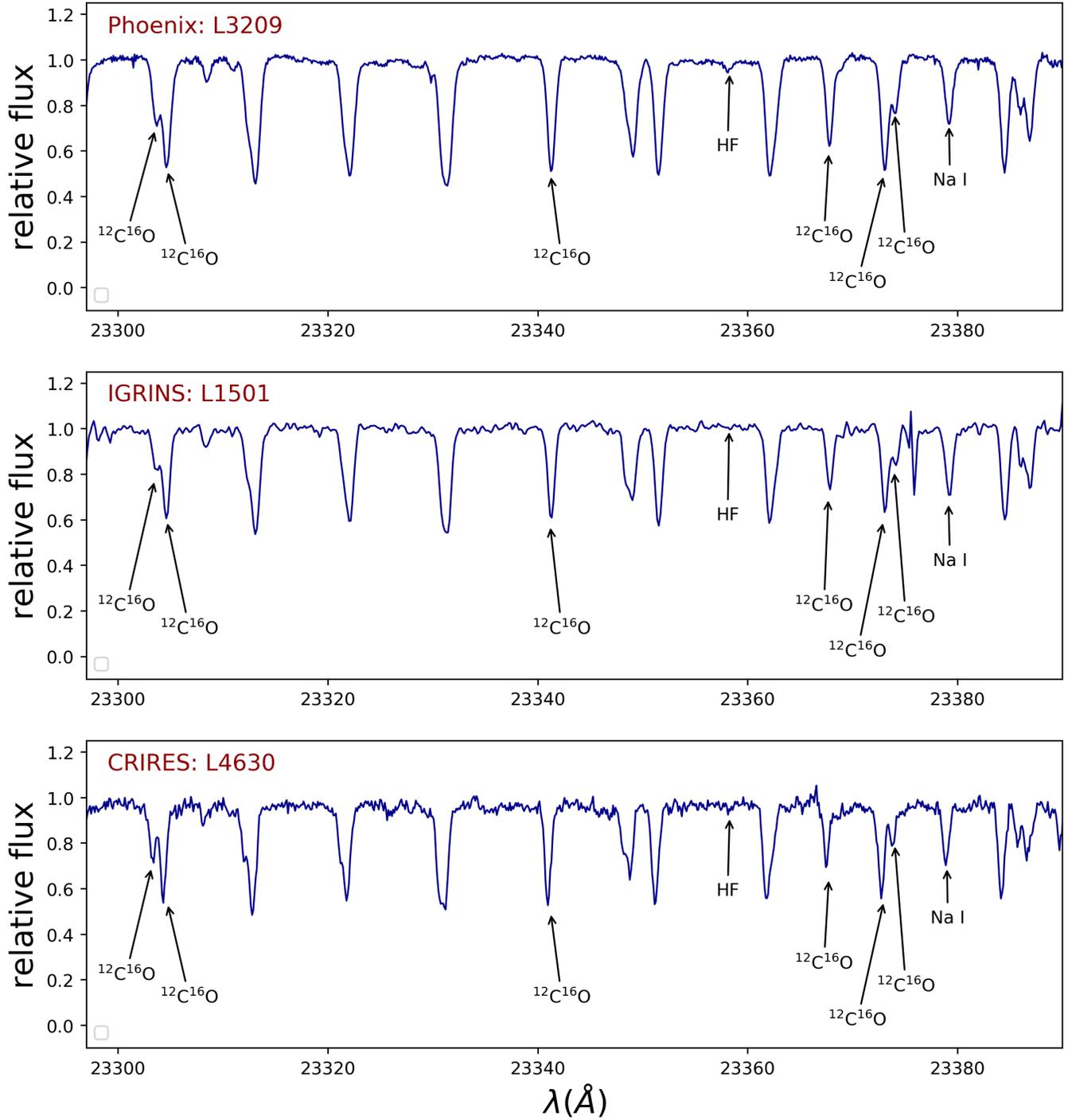}
   \caption{Top panel: The spectrum of the M4 star L3209 obtained with the Phoenix spectrograph (R = $\lambda / \Delta \lambda$ = 50,000). Middle panel: The spectrum of the M4 star L1501 obtained with the IGRINS spectrograph (R = $\lambda / \Delta \lambda$ = 45,000). Bottom Panel: The spectrum of the M4 star L4630 obtained with the CRIRES spectrograph (R = $\lambda / \Delta \lambda$ = 100,000). The spectral lines of $^{12}$C$^{16}$O, Na I and HF that were used to measure the abundances of carbon, sodium and fluorine, respectively are identified.}
   \label{fig:observed}
\end{figure}

The target stars are cool red giants (T$_{\rm eff}$ $\sim$ 3700 - 4500 K) belonging to the globular clusters M 4 and $\omega$ Centauri.
High-resolution high signal-to-noise infrared spectra in the K-band (including the region near $\lambda$2.3$\mu$m) were obtained using three different telescopes and spectrographs: the 8.1 m Gemini South Telescope with the NOAO Phoenix Spectrograph (Hinkle et al. 2003); the 2.7 m Harlan J. Smith Telescope at McDonald Observatory with the Immersion Grating Infrared Spectrometer (IGRINS; Yuk et al. 2010, Park et al. 2014); and the ESO 8.2 m Very Large Telescope and the CRyogenic high-resolution InfraRed Echelle Spectrograph (CRIRES; K\"ufl et al. 2004).
Table \ref{tab:observations} lists the observed targets along with the corresponding spectrograph used in each observation; two stars (L1514 and L3209) were observed using both the Phoenix and the IGRINS spectrographs. Figure \ref{fig:observed} shows examples of reduced spectra for three sample stars, each of them observed with one of the spectrographs: Phoenix (top panel), IGRINS (middle panel) and CRIRES (bottom panel). 

The Gemini / Phoenix spectra are single-order echelle spectra and the observations were taken at a resolution R = $\lambda/\Delta\lambda$ = 50,000.
The wavelength region shown in the different panels of Figure \ref{fig:observed}, between $\sim$23,300 -- 23,400 \AA, corresponds to the entire wavelength coverage of the observed Phoenix spectra and this spectral region has been selected by Cunha et al. (2003) and Smith et al. (2005) as it contains
the molecular HF (1-0) R9 line, as well as a number of CO lines from the 2-0 and 3-1 vibration bands, and a Na I doublet (Figure \ref{fig:observed}): one of the Na I lines is heavily blended with a CO line, so only one of the Na I lines was used to determine Na abundances. 
The Phoenix spectra were reduced to one dimension using standard IRAF routines in the previous studies by Cunha et al. (2003) and Smith et al. (2005).
IGRINS is a cross-dispersed near-infrared spectrograph covering the entire H and K bands; the observations were taken at a resolution R = $\lambda$/$\Delta\lambda$ = 45,000. The IGRINS spectra have been extracted to 1D and wavelength calibrated using various routines available within the "imred" and "echelle" packages of the standard spectral reduction software IRAF.
The VLT / CRIRES spectra were collected and reduced in the previous study by de Laverny \& Recio-Blanco (2013a). The CRIRES spectra have a resolution R = $\lambda/\Delta\lambda$ = 100,000 and the spectra were reduced with the ESO / CRIRES pipeline, and standard IRAF routines were used to remove the telluric contributions.

\section{Stellar Parameters}

The stellar parameters and metallicities for the studied stars come from previous works in the literature.
Table \ref{tab:observations} presents the effective temperatures (T$\rm_{eff}$), surface gravities  ($\log$ g),  microturbulent velocities ($\xi$) and metallicities ([Fe/H]) adopted for the stars.

The T$\rm_{eff}$, log g, and metallicities for the M 4 targets in this study, as well as in the previous study by Smith et al. (2005), were taken from Ivans et al. (1999), who based their T$\rm_{eff}$-scale on line-depth ratios (Gray 1994). 
The surface gravities in Ivans et al. (1999) were fixed by enforcing ionization equilibrium for the abundances of  Ti I / Ti II and Fe I / Fe II.
For the star L4404 the stellar parameters are from Suntzeff \& Smith (1991), where effective temperatures were obtained using (B--V) colors and log g values were determined using luminosities from Frogel et al. (1983) and an assumed red giant mass of 0.8M$_{\odot}$.
For the star L4630, values of T$\rm_{eff}$, log g and [Fe/H] were taken from de Laverny \& Recio-Blanco (2013a). Their T$\rm_{eff}$ was computed  from the (V −- K) and (V −- J) colors and the calibration of Ram\'irez \& Mel\'endez (2005). The $\log$ g values were calculated from fundamental relations involving T$\rm_{eff}$, bolometric corrections, and adopting a distance for M4 from Harris (1996).

The T$\rm_{eff}$, log g, and metallicities for the $\omega$ Centauri stars ROA 213, ROA 219 and ROA 324 were determined in Smith et al. (2000). 
This study used optical Fe I and Fe II lines to obtain the metallicities, (V -- K) and (B -- V) colors and the calibration from Ridgway et al. (1980) to derive the effective temperatures, and  fundamental relations 
to obtain surface gravities for the stars.
The stellar parameters and metallicities for ROA 43, ROA 132 and ROA 139 were taken from Norris \& Da Costa (1995) and Persson et al. (1980), with T$\rm_{eff}$s computed from (V -− K) colors, log g values computed from fundamental relations, and metallicities from equivalent width measurements of optical Fe I lines.

\section{Abundance Analysis}

The stellar atmospheric models used in this work are spherical models from the MARCS grid (Gustafsson et al. 2008; \url{http://marcs.astro.uu.se/}). For the adopted values of T$\rm _{eff}$, log g and metallicity for each star (Table \ref{tab:observations}), we interpolated their corresponding model atmospheres using the OSMARCS model grid computed for a mass of 1M$_\odot$, $\xi$ = 2.0 km$\cdot$s$^{-1}$, and {\it Standard} chemical composition (standard composition corresponds to $\alpha$-enhanced models for low metallicities). 

The C, Na and F abundances were determined by calculating synthetic spectra in Local Thermal Equilibrium (LTE) and using the code Turbospectrum (TS; Alvarez \& Plez 1998; Plez 2012), noting that TS is compatible with the adopted spherical model atmospheres given its spherical radiative transfer treatment.
Once the synthetic spectra were computed, the syntheses were manipulated using the plotting and broadening routines from the MOOG code (Sneden 1973) and 
best-fit abundances were derived by minimizing the differences between observed and synthetic spectra across the line profile.
For all stars, the model spectra were broadened by adjusting the full width at half-maximum (FWHM) of a single Gaussian that corresponds to the convolution of the instrumental profile, macroturbulent velocity, as well as, any contributions from v$\cdot$sini, although these should be insignificant given that the stars being analysed are red giants.

The microturbulent velocities adopted for the M4 targets 
are from Ivans et al. (1999), and de Laverny \& Recio-Blanco (2013a). 
We also estimated microturbulent velocities from the empirical relation in Pilachowski et al. (1996) and from an empirical relation derived from the $\xi$ and log g results obtained by Alves-Brito et al. (2010). In a few cases we adjusted $\xi$ based on the analyzed CO lines, as discussed below.

\begin{figure}[t!]
  \centering
  \includegraphics[width=\textwidth]{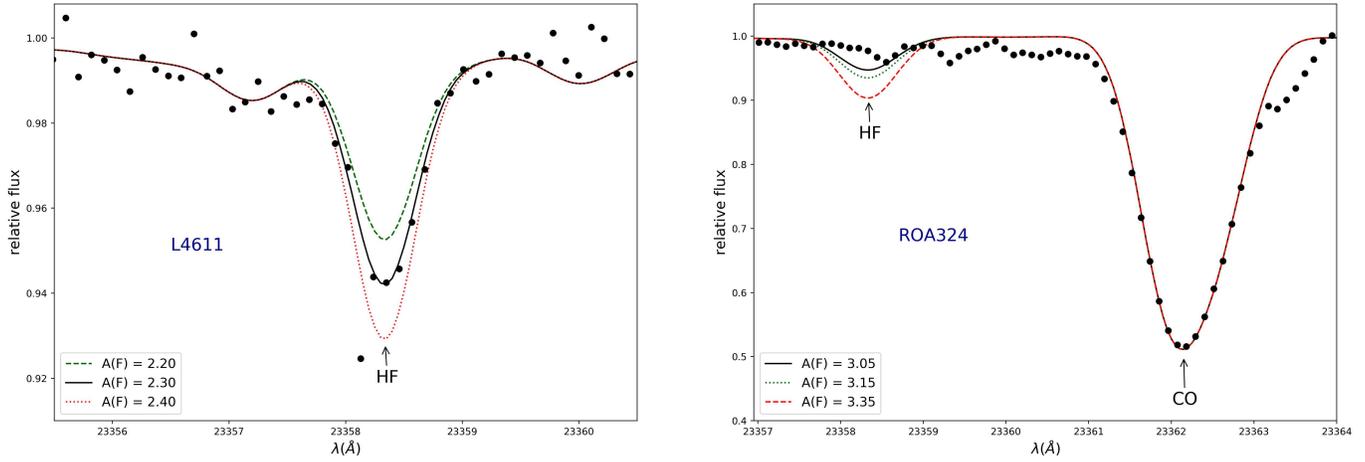}
  \caption{Left Panel: The observed and synthetic spectra of the M 4 star L4611. The synthetic spectra were computed for three different fluorine abundances: A(F) = 2.20 (green dashed line), 2.30 (best synthesis; solid line) and 2.40 (red dashed line). 
  Right Panel: The observed and synthetic spectra of the $\omega$ Centauri star ROA 324. For this an upper limit fluorine abundance has been derived. The synthetic spectra shown were computed for A(F) = 3.05 (upper limit abundance; solid line), 3.15 and 3.35.}
  \label{fig:synthesis}
\end{figure}

Table \ref{tab:lines} presents the spectral lines used in the abundance determinations, along with the atomic and molecular data (excitation potential ($\chi$), log $gf$, and dissociation energy (D$_o$)) adopted for each transition.
To keep the abundance analysis as homogeneous as possible, we have restricted the line selection to the wavelength range between $\sim$ 23,300 -- 23,390 \AA, which is the range covered by the spectra obtained with the Phoenix spectrograph (shown in Figure \ref{fig:observed}).
The fluorine abundances were derived from spectrum synthesis of the HF(1–-0) R9 rotational-vibrational line at 23,358.329 \AA. 
As discussed previously, the excitation potential used for this molecular transition has been updated from 0.480 eV (Jorissen et al. 1992) to 0.227 eV (Decin 2000; J\"onsson et al. 2014a,b) and its dissociation energy from 5.820 eV (Cunha et al. 2003) to 5.869 eV (Sauval \& Tatum 1984, J\"onsson et al. 2014a). 
Synthetic profiles for the HF line for the sample stars L4611 from M 4 and ROA 219 from $\omega$ Centauri are shown in Figure \ref{fig:synthesis}. 
These are the syntheses that best fit the observed spectra for the HF(1--0) R9 lines.
To determine sodium abundances, one of the Na I doublet lines was used, consisting of two closely spaced components at 23378.945 \AA \ and 23379.139 \AA. 

The $^{12}$C abundances were determined using six $^{12}$C$^{16}$O lines; these are the same CO lines analyzed in the previous study by Smith et al. (2005). 
CO is an important molecule in the molecular equilibrium of cool giant atmospheres, as are, although to a lesser degree, the OH and CN molecules (also with measurable transitions in the near-infrared).
To derive the carbon abundances for the target stars, we adopted their O and N abundances from the studies of Ivans et al. (1999) and Smith et al. (2005) (for M4 stars); and Norris \& da Costa (1995) and Smith et al. (2000) (for $\omega$ Centauri stars). 
The CO lines at $\sim$ $\lambda$2.3$\mu$m are also useful to constrain the microtubulent velocity as the strongest lines are sensitive to the microturbulent velocity parameter. As mentioned above, the sample CO lines were used in order to test the adopted microturbulent velocities for the studied stars and good agreement was obtained. 

Table 3 presents the abundance results for all measured lines of CO, the Na I line, and the (1--0)R9 HF line; the mean abundance values for carbon and standard deviations of the mean are also listed for all studied stars.
In addition, we also present abundance results for the benchmark star Arcturus from an analysis of an FTS spectrum (Hinkle, Wallace \& Livingston 1995) and adopting stellar parameters T$\rm _{eff}$= 4275 K; log g = 1.70 (Smith et al. 2013); micro = 2.0 km s$^{-1}$ and [Fe/H]= -0.54 (Guer\c{c}o et al. in preparation).

\subsection{Abundance Uncertainties}

The uncertainties in the derived F, C and Na abundances due to the errors in the stellar parameters can be estimated by establishing the sensitivity of these abundances to changes in the stellar parameters corresponding to their typical errors. We estimate that the uncertainties in the stellar parameters are: $\delta$T$\rm_{eff}$ = +100 K, $\delta$log g = +0.25 dex, $\delta$[Fe/H] = +0.1, and $\delta \xi$ = 0.3 km$\cdot$s$^{-1}$.
We used the model atmosphere for the star L4611 (T$\rm_{eff}$ = 3725 K, log g = 0.30, [Fe/H]= -1.11, and $\xi$ = 1.70 km$\cdot$s$^{-1}$) as baseline and changed the parameters by the estimated uncertainties; 
the variations in the abundances are in Table \ref{tab:disturbance}. 
The estimated error ($\Delta$) for each element is the sqrt of the sum in quadrature of the corresponding change in abundance for $\delta$T$\rm_{eff}$, $\delta$log g, $\delta$[Fe/H] and $\delta \xi$. 
The fluorine abundances are most sensitive to the effective temperature, with errors in T$\rm_{eff}$ accounting for most of the F abundance uncertainties. The sodium abundances are also sensitive, but not as much, to the effective temperature, while the derived carbon abundances show more sensitivity to the microturbulent velocity and to a lesser degree to the metallicity of the model atmosphere.

\section{Results}

\subsection{Abundance Trends with Stellar Parameters}

As noted previously, the stellar parameters for most of the M 4 red giants were taken from Ivans et al. (1999) and, due to the sensitivity of the HF line strength to the effective temperature, the derived fluorine abundances are plotted versus T$_{\rm eff}$ in the left panel of Figure \ref{fig:Teff_logg_F_M4}; the stars are coded as ''CN strong'' (filled red circles) and ''CN weak'' (filled blue circles) according to Ivans et al. (1999).  Included in Figure 3 are the results from Smith et al. (2005) as an illustration of the change in the F abundance that results from the change in the HF excitation energy (from $\chi$=0.48 eV to 0.227 eV).  In both the previous results from Smith et al. (2005) and the updated F abundances presented here, there is a trend of A(F) with T$_{\rm eff}$, which is not expected, as low-mass red giants (M$\sim$0.85M$_{\odot}$), such as those in M4, do not deplete significantly F during mixing on the RGB 
(e.g., stellar models from Lagarde et al. 2012).
The right panel of Figure 3 also shows A(F), but here as a function of log g, where A(F) is found to decrease with decreasing log g.  In a globular cluster, T$_{\rm eff}$ and log g are correlated as stars evolve up the RGB to both lower values of T$_{\rm eff}$ and log g, so the two trends in Figure 3 are consistent.  As can be seen in Table 4, the fluorine abundance derived from HF is more sensitive to likely errors in the effective temperature scale when compared to such errors in log g.  

In order to check a possible systematic error in the T$_{\rm eff}$-scale from Ivans et al. (1999), the Na abundances derived here are plotted against both effective temperature and log g in the left and right panels, respectively, of Figure 4.  The points are plotted as in Figure 3, as either CN strong or CN weak and the large scatter in the Na abundances in M4 is real, with the CN-strong and CN-weak stars segregating as separate sequences, as discussed in Ivans et al. (1999).  There is no underlying significant trend of the derived Na abundance with T$_{\rm eff}$ or surface gravity; a linear fit to the Na abundances versus T$_{\rm eff}$ finds no significant slope, with a change of only 0.05 dex from a linear fit from 3750 K to 4650 K.
The sodium abundance is less sensitive to both the effective temperature and surface gravity than the fluorine abundances derived from HF, so a small systematic trend in the Ivans et al. (1999) might still exist to produce a trend of A(F) with T$_{\rm eff}$.

In order to check whether an independent T$_{\rm eff}$ scale would also produce a significant trend in A(F) with T$_{\rm eff}$, a set of effective temperatures were derived for the M 4 red giants using the near-IR color (J-K$_{\rm s}$) with photometric effective temperature relations from both Gonz\'alez Hern\'andez \& Bonifacio (2009), and Bessell, Castelli \& Plez (1998).  

Non-negligible interstellar reddening exists along the line-of-sight to M4, which must be taken into account in deriving photometric effective temperatures (this is one of the reasons that Ivans et al. (1999) relied on a line-depth ratio T$_{\rm eff}$-scale).  
Ivans et al. (1999) mapped the reddening across M 4 and found variations around a mean of E(B-V)=0.32 $\pm$ 0.05.  Clayton \& Mathis (1988) find that E(J-K) = 0.555 E(B-V) and we corrected all observed (J-K$_{\rm s}$) values for reddening using two methods: in one case the mean value of E(J-K$_{\rm s}$)=0.18$\pm$0.03 for M 4 was used, while in the other, the individual values of reddening, as derived by Ivans et al. (1999; their Table 4), were applied.  A comparison of the effective temperatures derived from using individual reddening values for each star, with those obtained from an average cluster reddening, results in a small mean offset and a scatter of about 60 K ($\Delta$T$_{\rm eff}$ = -20 K $\pm$ 61 K, in the sense of derived T$_{\rm eff}$s using 'individual reddening' minus 'mean reddening').  

A scatter of $\sim$60 K is the expected value from a scatter of 0.03 in (J-K$_{\rm s}$).  
Since the offset and scatter within the T$_{\rm eff}$ values derived from using a mean reddening or using the individual reddening values are small, the photometric effective temperatures derived from the mean reddening to M 4 are compared to those from Ivans et al. (1999); the photometric T$_{\rm eff}$'s from (J-K$_{\rm s}$) are cooler than those derived from the line-depth method from Ivans et al. (1999), with the mean $\Delta$T$_{\rm eff}$(line depth - (J-K)) = +118 $\pm$ 45 K.  There is no slope in the $\Delta$T$_{\rm eff}$ - T$_{\rm eff}$ points, so adopting this photometric T$_{\rm eff}$-scale will not remove the trend in A(F) versus T$_{\rm eff}$ presented in Figure 3. 
Although adopting the photometric scale for effective temperatures would not remove the trend of A(F) with T$_{\rm eff}$, it would systematicaly lower the absolute fluorine abundances by $\sim$ 0.2 dex.
This trend is likely due to some other effect which we investigate below.

As mentioned in the introduction, two other studies have derived fluorine abundances in globular cluster red giants: Yong et al. (2008) for NGC 6712 and de Laverny \& Recio-Blanco (2013a) for 47 Tuc.  Both of these globular clusters have metallicities that are less metal-poor but not too different from that of M 4 (with 47 Tuc having [Fe/H] $\sim$ -0.75 and NGC 6712 having [Fe/H] $\sim$ -1.0) and a plot of their fluorine abundances also reveals a trend of A(F) decreasing with decreasing T$_{\rm eff}$, having a slope similar to that found by us for M 4.  
Both Yong et al. (2008) and de Laverny \& Recio-Blanco (2013a) use their own independent methods of deriving effective temperatures, so it is unlikely that the T$_{\rm eff}$ trend in all three studies is caused by an incorrect effective temperature scale.  
We speculate that this trend found in fluorine abundances derived from HF is due to a physical effect related to the spectroscopic analysis of low metallicity red giants.  All globular cluster fluorine abundances show this trend and the effect caused by the slope with T$_{\rm eff}$ can be corrected in order to compare relative inter-cluster abundances.

A physical cause of the A(F) - T$_{\rm eff}$ trend may be due to the use of 1D static model atmospheres, as discussed by Li et al. (2013) in their analysis of HF in metal-poor field red giants.  Li et al. (2013) did test syntheses of HF using convective 3D model atmospheres from the grid of Ludwig et al. (2009) and found significant differences between 1D static and 3D hydrodynamic modelling of the HF(1-0) R9 line.  They present corrections for a series of red giant model atmospheres at a metallicity of [m/H]=-2.0 with [$\alpha$/m]=+0.4 and these corrections follow a similar trend with T$_{\rm eff}$ as the observed globular cluster red giants.  The 3D results from Li et al. (2013) suggest that the A(F)-T$_{\rm eff}$ trends found in the 1D static model atmosphere analyses of globular cluster red giants by Yong et al. (2008), de Laverny \& Recio-Blanco (2013a), and this study may result from 3D hydrodynamic effects, which can be removed empirically as a first order (linear) correction, which is plotted as the straight line in the left panel of Figure 3, and this will be discussed in Section 6.2.2. For reference, the fluorine abundances corrected for the trend with T$_{\rm eff}$ are presented in the last column of Table 3.

Figure \ref{fig:carbon_Mbol} presents the carbon abundances versus bolometric magnitude (M$_{bol}$) for the M4 stars in our sample. 
Overall, the results indicate a trend of decreasing $^{12}$C abundances with increasing red giant luminosity (a decrease in M$_{bol}$ implies an increase in luminosity), or, lower T$_{eff}$s.
Such a trend was also observed in previous studies of M 4 and other globular clusters (e.g., Langer et al. 1986; Suntzeff \& Smith 1991; Bellman et al. 2001; Smith et al 2005; Kirby et al. 2015). 
As done previously, we segregate the M 4 sample into the CN-strong (carbon-poor nitrogen-rich stars; red circles) and CN-weak (stars with more pristine values of carbon and nitrogen; blue circles) populations. 
Two sequences are roughly defined given the segregation in carbon abundances of the CN-weak (carbon 'rich') and CN-strong (carbon poor) stars and within a sequence we see the effect of internal mixing occurring as the star ascends the RGB.
The observed decrease in $^{12}$C as a low-mass giant ascends the RGB is not surprising, as recent stellar models that include rotational and thermohaline mixing predict such a trend (Lagarde et al. 2012).

\subsection{Comparisons with Abundances from the Literature}

One relevant result from this study is that the derived fluorine abundances are, as expected, significantly lower than those obtained in our previous studies from the literature, which adopted inconsistent values for the excitation potential ($\chi$) and dissociation energy for the analyzed R9 HF transition. 
The offsets between our derived fluorine abundances for M 4 and those from Smith et al. (2005; shown as open circles) and de Laverny \& Recio-Blanco (2013a; shown as the open triangle) can be seen in Figure 3. 

Figure \ref{fig:carbon} (left panel) shows the comparison between our derived carbon abundances and those determined in different studies in the literature.
The comparison between the carbon abundances derived here and those in Smith et al. (2005) shows good agreement (a mean difference of $<$ $\delta$A(C)$>$ = - 0.10 $\pm$ 0.11). 
Smith et al. (2005) determined carbon abundances using the same $^{12}$C$^{16}$O lines at 2.3 $\mu$m that we analyze here (Table \ref{tab:lines}).
On the other hand, when compared to $^{12}$C results obtained from the analysis of low-resolution spectra at 2.2 $\mu$m by Suntzeff \& Smith (1991) there is a much larger discrepancy: the mean carbon abundance in this study is systematically larger than theirs by $<$ $\delta$A(C)$>$= 0.56 $\pm$ 0.28. Such a difference comes, in part, from inadequate microturbulent velocities adopted in Suntzeff \& Smith (1991; see discussion in Smith et al. 2005).
When compared to the carbon abundances from the high-resolution optical study by Norris \& da Costa (1995) there is a very small offset in the mean, but a larger abundance scatter ($<\delta$A(C)$>$ = -0.03 $\pm$ 0.25);  the difference with the results from low-resolution spectra (R = $\lambda/\Delta \lambda$ = 600; 1200) from Stanford et al. (2010) shows a larger systematic offset: 
$<$ $\delta$A(C) $>$ = 0.33 $\pm$ 0.22.

Figure \ref{fig:carbon} (right panel) shows the comparison between our derived sodium abundances and those determined in the literature.
For the globular cluster M 4, the sodium abundances that we determine are in a good agreement with the results from the high-resolution optical study by Ivans et al. (1999): $<\delta$A(Na)$>$ = -0.01 $\pm$ 0.04;  with results from the infrared study by de Laverny \& Recio-Blanco (2013a): $<\delta$A(Na)$>$ = +0.11 $\pm$ 0.11 and with results from MacLean et al. (2018): $<\delta$A(Na)$>$ = +0.04 $\pm$ 0.13. 
Our sodium abundances for $\omega$ Centauri are in agreement with the optical high-resolution study by Smith et al. (2000) having an average difference of $< \delta$A(Na)$>$ = 0.03 $\pm$ 0.06. 
Other works in the literature have systematically higher sodium values than ours, presenting differences of $< \delta$A(Na)$>$ = - 0.38 $\pm$ 0.18 for Norris \& da Costa (1995), $< \delta$A(Na)$>$ = - 0.22 $\pm$ 0.06 for Johnson \& Pilachowski (2010), $< \delta$A(Na)$>$ = - 0.48 $\pm$ 0.26 for Stanford et al. (2010; R = 600 spectra), and $< \delta$A(Na)$>$ = - 0.27 $\pm$ 0.42 for Stanford et al. (2010; R = 1,200 spectra).

\section{Discussion}

\subsection{C, O and Na Abundance Patterns}

Globular clusters are now known to have more than one stellar generation and not to be simple stellar populations (Gratton et al. 2012; Marino 2018).  
As discussed in the introduction, stars from different generations in a globular cluster show the distinct pattern (not seen in the general field star population) of depletion in the C, O, and Mg abundances corresponding to enhancements in the N, Na, and Al abundances. These abundance variations in globular clusters, or, anti-correlations in the abundances of C-N, Na-O, Mg-Al are the result of changes in the chemical abundances due to hydrogen burning (proton captures) occurring in the CNO, NeNa and MgAl cycles.

\subsubsection*{M 4}

The left panel of Figure \ref{fig:C_O_Na_M4} presents the [Na/Fe] versus [O/Fe] abundances for the studied M 4 targets. 
Our results confirm the anti-correlation between O and Na that has been previously found for M 4 by Ivans et al. (1999), Smith et al. (2005), Carretta et al. (2009), and Marino et al. (2011).
Such an anti-correlation pattern for M 4 is consistent with similar results found in the literature for other galactic globular clusters (e.g., Ramirez \& Cohen 2002 for M 71, Carretta et al. 2004 for 47 Tucanae, Carreta et al. 2009 for 17 globular clusters and Carretta et al. 2014 for NGC 4833).
In fact, the anti-correlation pattern between Na-O abundances is observed in virtually all globular clusters in the Milky Way and this abundance anti-correlation is inferred as the chemical signature that 'tags' a globular cluster population  (Carretta et al. 2009; Marino 2018).

An anti-correlation between the abundances of Na and C for M 4 is also obtained in this study and this is shown in the right panel of Figure \ref{fig:C_O_Na_M4}.
The observed anti-correlation between Na and C can be understood as the result of the decrease in the carbon abundances due to the CN cycle (e.g, Sweigart \& Mengel 1979; Marino et al. 2016), and the increase in the Na abundances as a result of the the NeNa cycle.

\subsubsection{$\omega$ Centauri}

The [O/Fe] and [C/Fe] versus [Na/Fe] abundances for the studied $\omega$ Centauri stars are presented in the top and bottom panels of Figure \ref{fig:O_vs_Na-C_vs_Na-o_Cen}, respectively. We also show for comparison, results from the optical by Smith et al. (2000), and results from the much larger optical study by Marino et al. (2012). 
Given the complexity in investigating abundance anti-correlations in a globular cluster having an extreme metallicity (iron abundance) spread, such as is observed for $\omega$ Centauri (Johnson \& Pilachowski 2010), we divide, as in Marino et al. (2012), the stars into the different stellar populations, corresponding to the following metallicity bins (five panels in Figure \ref{fig:O_vs_Na-C_vs_Na-o_Cen}): [Fe/H] $<$ -1.90 dex; -1.90 dex $\leq$ [Fe/H] $<$ -1.65 dex; -1.65 dex $\leq$ [Fe/H] $<$ -1.50 dex; -1.50 dex $\leq$ [Fe/H] $\leq$ -1.05 dex; and [Fe/H] $>$ -1.05 dex. 

We can see from the results shown in the top panels of Figure \ref{fig:O_vs_Na-C_vs_Na-o_Cen} that the overall pattern of Na-O anti-correlation holds for $\omega$ Centauri, or, stars richer in sodium tend to have lower oxygen abundances and vice-versa. A similar result is also seen for [Na/Fe] vs [C/Fe] (bottom panels).
The pattern shown in the different panels might be suggestive of a progression in the anti-correlation slope, which would seem to be progressively displaced towards higher values of [O/Fe], (less clear for carbon) as the metallicity increases, until reaching the most metal-rich population ([Fe/H] $>$ -1.05 dex), which does not show a clear Na-O or C-O anti-correlation pattern, having stars with the highest sodium, and oxygen abundances, as well as carbon. 
The results for the few $\omega$ Centauri stars in our study (blue circles), together with those from Smith et al. (2000; green circles), follow the general pattern delineated by the much larger sample by Marino et al. (2012; Xs) for the different $\omega$ Centauri populations.

\subsection{Fluorine Abundance Patterns}

\subsubsection{Previous Fluorine Studies in Globular Clusters}

The literature on chemical abundance patterns in globular clusters is quite extensive and rich but the number of studies that focused on fluorine abundance determinations is much more limited.
Some of the fluorine results obtained for stars in globular clusters are discussed below.

One of the first works to study fluorine in a globular cluster was Smith et al. (2005). 
Their fluorine abundances, derived using the HF R9 line, indicated that there are fluorine abundance variations in M 4, which are anti-correlated with the Na abundances. 
More recently, de Laverny \& Recio-Blanco (2013a) also studied fluorine in M 4 obtaining a HF measurement for one additional star and two upper limit fluorine abundances. Their results were in agreement with the conclusions from Smith et al. (2005) that the abundances of fluorine and sodium are anti-correlated. 
As mentioned before, the excitation potential of the HF R9 line used in these studies and resulting fluorine abundances are being revised here.

The globular cluster NGC 6712, with a similar metallicity to M 4, was studied by Yong et al.(2008), who found significant fluorine abundance variations, with two of their targets being fluorine-poor and three of them fluorine-rich; the measured values of A(F) showed significant scatter ($\sim$ 0.6 dex).
(We note that this study also used the inconsistent excitation potential for the HF R9 transition).
For the most metal-rich globular cluster 47 Tuc, de Laverny \& Recio-Blanco (2013a) were able to measure the HF R9 line and obtained a variation of $\sim$ 0.4 dex in A(F) and a rather similar one in A(Na). Unlike for M 4, an anti-correlation between the F and Na abundances for 47 Tuc was not found. 

For the globular cluster M 22, there are three studies in the literature with somewhat conflicting results; the lack of consensus in the fluorine abundance results for M 22 ultimately highlights the difficulty in analyzing the weak HF R9 line within telluric contamination.
Alves-Brito et al. (2012) targeted 11 red giant stars in M 22 and reported fluorine abundances spanning a range of $\Delta$A(F)$\sim$ 0.6 dex, which is comparable to the abundance spread also observed for other light elements in M 22. 
Two stars in their sample seem to follow the general trend of [F/Fe] increasing with [O/Fe] and decreasing with [Na/Fe], although the most fluorine-rich stars in their sample do not follow this general trend.
Independent determinations of fluorine abundances in M 22 were made by D'Orazi et al. (2013). This study selected three stars from the more metal-poor M 22 population and three stars from the more metal-rich one and within these two populations, O-rich (Na-poor) and O-poor (Na-rich) stars were selected. They found a correlation between F and O abundances, and an anti-correlation between the abundances of F and Na. These F abundance results for M 22 were questioned by de Laverny \& Recio-Blanco (2013b) from a re-examination of the same CRIRES spectra previously analyzed in D'Orazi et al. (2013). de Laverny \& Recio-Blanco (2013b) concluded that the fluorine abundance determinations for M 22 were doubtful or represented most likely non-constraining upper limits.

\subsubsection{Fluorine Abundance in M 4 and $\omega$ Centauri}

As discussed in Section 6.1, our results indicate that the abundances of the elements C, O, and Na are not constant in the globular cluster M 4, but rather show anti-correlations between Na-O, and Na-C, as would be expected to result from the CNO and Ne-Na hydrogen-burning cycles operating in the first stellar generation in M 4, with a second stellar generation forming from gas that was polluted by the first generation. 

Our results also indicate that there is an anti-correlation between F and Na in M 4 (Figure \ref{fig:F_O_Na_M4}, left panel); the trend is similar to that obtained previously by Smith et al. (2005) and de Laverny \& Recio-Blanco (2013a), although the revised absolute [F/Fe] values are lower than previously found in those earlier studies due to the lower excitation potential of the HF line.

\begin{figure}[!htb]
   \centering
   \includegraphics[width=1.0\textwidth]{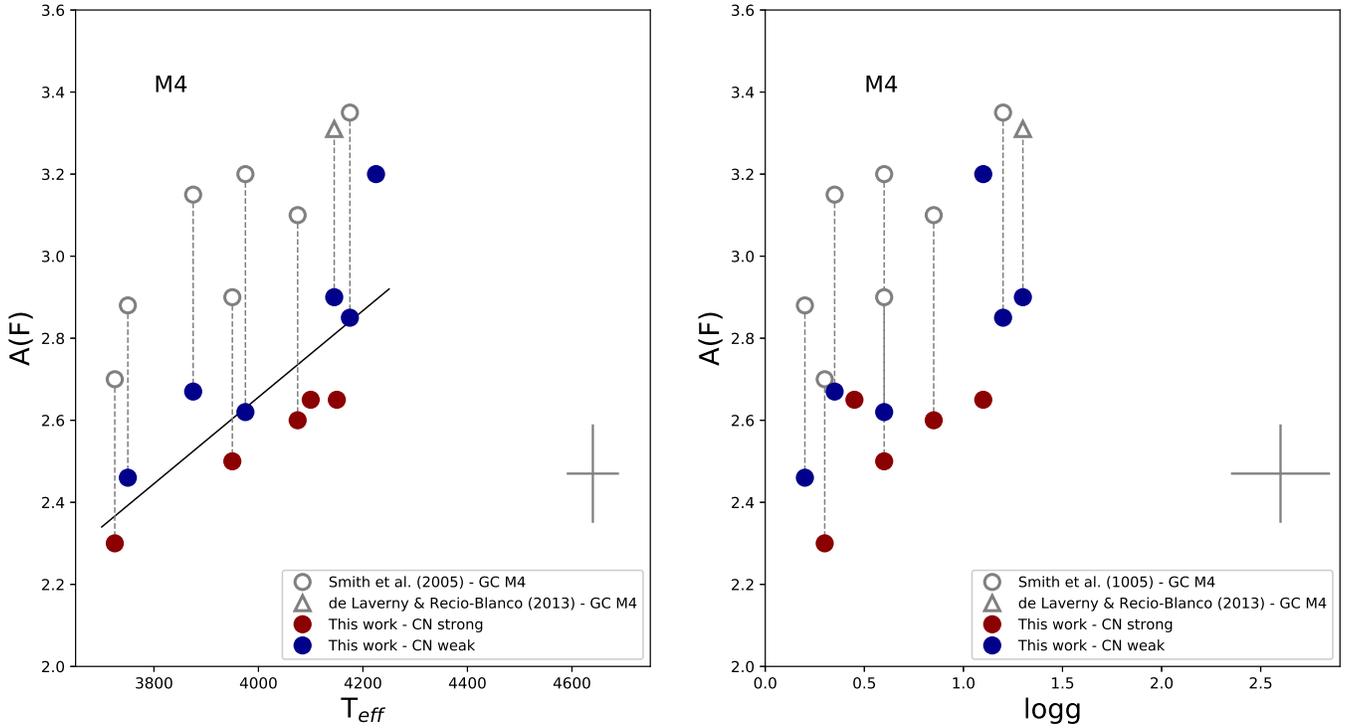}
   \caption{Fluorine abundances for the M4 red giants analyzed in this work are presented versus T$_{eff}$ (left panel) and log g (right panel). 
   The filled blue circles represent the CN-weak stars and the filled red circles represent the CN-strong stars, as classified in Ivans et al. 1999).
   The previous results from Smith et al. (2005; open grey circles) and de Laverny \& Recio-Blanco (2013; open grey triangle) are shown for comparison; the dashed lines connect the stars in common. 
   Upper limit fluorine abundances derived for three of the targets analyzed in de Laverny \& Recio Blanco (2013a) are not shown in the figure, given their higher effective temperatures ($T_{eff}$ $>$ 4300 K and log $>$ 1.5) and much higher fluorine upper limits. Typical error bars are shown; these correspond to the estimated internal uncertainties in the effective temperatures of Ivans et al. (1999) and an uncertainty in log g = 0.25 dex.}
   \label{fig:Teff_logg_F_M4}
\end{figure}

The Na-F abundance anti-correlation can be understood in a simple way, with a primordial stellar population in M 4 having destroyed fluorine in stars that were massive enough to reach interior temperatures high enough for the NeNa cycle to occur, as evidenced by the observed anti-correlation between Na and O in M 4. In this simple picture, a fraction of initial $^{19}$F in the primordial cluster material was consumed by (p,$\gamma$) and (p,$\alpha$) reactions, while $^{23}$Na was synthesized by the Ne-Na cycle; this leads to the anti-correlation between A(F) and A(Na), between the primordial (first generation) and second generation of stars.  
In Section 5.1 the decline in fluorine abundance with decreasing T$_{\rm eff}$ was discussed as a possible manifestation of inadequacies of 1D modelling of the very temperature sensitive HF lines, with a fitted straight line shown in the left panel of Figure \ref{fig:Teff_logg_F_M4} as an illustration of this trend.  Even with the T$_{\rm eff}$ trend in A(F), it should be noted from Figure \ref{fig:Teff_logg_F_M4} that the CN-strong and -weak stars segregate on either side of this line, with the CN-strong stars falling systematically below the line and the CN-weak stars falling above the line.  Because of this difference in fluorine abundance between the two families of stars, the Na-F anti-correlation from Figure \ref{fig:F_O_Na_M4} (left panel) is still apparent.  The A(F)-T$_{\rm eff}$ trend from Figure \ref{fig:Teff_logg_F_M4} can be removed from the measured F abundances and these ''corrected'' F abundances are used in the right panel of Figure \ref{fig:F_O_Na_M4}; in this case, the Na-F anti-correlation becomes smoother and displays less scatter.  The result of these tests provides confidence that a real Na-F abundance anti-correlation exists in the two stellar families in M 4.

Our results for $\omega$ Centauri are less conclusive given the small number of red giants analyzed and the fact that the fluorine results for the $\omega$ Centauri sample are, for the most part, upper limit fluorine abundances. 
The only $\omega$ Centauri star with a measured fluorine abundance is ROA 219 (T$_{\rm eff}$ = 3900 K; log g = 0.70).
This star has a metallicity [Fe/H]= -1.20 and belongs to the intermediate metallicity population in $\omega$ Centauri (-1.50 dex $\leq$ [Fe/H] $\leq$ -1.05 dex). 
Having a metallicity that is very close to that of M 4, this star may be compared to the M 4 results shown in Figure \ref{fig:F_O_Na_M4}. The fluorine and sodium abundances derived for ROA 219 ([F/Fe]$\sim$-0.4; [Na/Fe]$\sim$-0.1) puts it close to, or, just slightly below, the star with lowest Na abundance in our M 4 sample (the latter belongs to the 'high fluorine - low sodium' group in M 4). Our $\omega$ Centauri sample has three other $\omega$ Centauri targets in the same metallicity bin and, interestingly, although they all have upper limit fluorine abundances, 
they also all have low A(Na) (lower than [Na/Fe] $\sim$-0.6), which would make them part of the primordial population. When considering their C and O abundances, their low fluorine abundances appear to be consistent with their loci in the Na - O and Na - C  panels in Figure \ref{fig:O_vs_Na-C_vs_Na-o_Cen} (filled blue symbols), that also would indicate a primordial origin as their oxygen and carbon abundances are high. 
Fluorine abundance upper limits are also derived for the two additional $\omega$ Centauri stars, ROA 213 and ROA 324; 
these upper limit abundances do not provide any additional constraints.

\subsection{Comparison of the M4 Fluorine and Sodium Abundances with Stellar Models}

Figure \ref{fig:F19vNa23All} provides a summary view and comparison of derived M4 red giant fluorine and sodium abundances with those arising from a sample of stellar models from both Lagarde et al. (2012) and Smith et al. (2005).  This figure plots the Na versus F abundances, with the M 4 red giants shown as filled circles and classified as either CN weak (blue), CN strong (red), or CN type unknown (green). These abundances are corrected for the T$_{\rm eff}$ trend discussed in Section 5.1.
Included in Figure \ref{fig:F19vNa23All} are stellar model results presented in Smith et al. (2005) for a 6.5M$_{\odot}$ hot bottom burning (HBB) AGB star; this model is of the same type as described in detail in Fenner et al. (2004) and Campbell \& Lattanzio (2008).  The open square at the high-Na and low-F part of the figure is the final yield from this HBB AGB model, while the continuous curve connects this yield to the initial F and Na abundances in the primordial M 4 stellar population (high-F and low-Na abundances); 
the curve connects the initial F and Na abundances in the model with the final HBB AGB abundances via different mixing fractions of the initial and final F and Na abundances.
The other model results presented in Figure 10 are from Lagarde et al. (2012) for masses of 4.0M$_{\odot}$ and 6.0M$_{\odot}$ and include models with standard dredge-up (St), along with models that include both rotational and thermohaline mixing (Rot-Th).  The abundances are the surface abundances that result from the various mixing episodes that occur as a result of RGB and AGB evolution. In the case of Lagarde et al. (2012), the models are followed to the tip of the AGB and end there at the cessation of He-burning and do not include the final stripping of the stellar envelope down to the remnant white dwarf core.  

Within the overlap of mixing episodes covered by both the Lagarde et al. (2012) and Smith et al. (2005) models, the comparison is good, with the Lagarde et al. models predicting slightly more F depletion for a given Na enhancement.  It is worth noting that the Lagarde et al. (2012) 6.0M$_{\odot}$ model, that includes rotational and thermohaline mixing, produces F along the upper AGB (where it becomes a carbon star), but the model is halted before it undergoes HBB, which would destroy $^{19}F$ and synthesize $^{23}$Na, convert $^{12}$C to $^{14}$N, driving the C/O ratio to less than 1.0 and turning the former carbon star into a luminous S-type AGB star (Wood et al. 1983);
it is the lower mass AGB stars that are net producers of fluorine, as they are not massive enough to undergo HBB.
Both sets of models predict a large Na enhancement, with only a small F depletion, until the increase in Na reaches $\sim$+1 dex before F undergoes significant (~$>$-0.3 dex) depletion.  The observed abundances in M 4 suggest a somewhat larger depletion of F at a rather modest increase in the Na abundance: the fluorine abundance drops by about 0.2 dex over a +0.5 dex increase in A(Na).  Lower-mass stellar models produce less F depletion, so pollution in the early M 4 environment from lower-mass stars is not likely.  The observed behavior of A(F) as a function of A(Na) and its comparison to models is similar to that found by Fenner et a. (2004) in the globular cluster NGC 6752 for Na and O; Fenner et al. (2004) found that the observed depletion in oxygen was larger at a given Na increase than that predicted by stellar models.  In M 4 it is not clear what is causing the discrepancy between observations and models for Na and F.

\section{Conclusions}

Fluorine abundances in M 4 red giants, derived using updated and consistent values for the excitation potential and dissociation energy for the R9 HF (1-0) transition, reveal an anti-correlation with the sodium abundances (as previously found by Smith et al. 2005).  A comparison of the observed F and Na abundances with those produced in stellar models by Lagarde et al. (2012) and Fenner et al. (2004, with predictions for fluorine presented in Smith et al. 2005) suggest that the observed level of F depletion ($\sim$0.2 dex) in the second stellar generation in M 4 is larger at a given Na abundance enhancement than that predicted by the models.  The difference in the F depletion between observed and model predictions is not large; an increase in the F depletion of $\sim$0.1 dex would lead to rough agreement between the models and the abundances derived in this study. Such difference could possibly be explained by either improvements to the abundance analyses or in the modelling.  Further observations of HF in globular clusters are certainly warranted.

The comparison between the models and abundance results generally indicates that the stars responsible for driving the fluorine and sodium abundance anti-correlation in M 4 are best explained by the more massive AGB stars (M $\ge$ 6M$_{\odot}$), as lower mass models produce even less fluorine depletion. If such massive stars were the main contributors to the chemical evolution in M 4,
it would suggest that the early M 4 environment was polluted on a timescale of $\le$10$^{8}$ years (the evolutionary timescale for a $\sim$6 M$_{\odot}$ star).

Our results for $\omega$ Centauri are less conclusive given the small number of targets analyzed and the fact that the fluorine results for the $\omega$ Centauri sample are, for the most part, upper limit fluorine abundances.

\acknowledgments
\section{Acknowledgments}

We thank the referee for suggestions that helped improve the paper. 
R. G. acknowledges support from CAPES fellowship at Observat\'orio Nacional - Rio de Janeiro.
H. J. acknowledges support from the Crafoord Foundation, Stiftelsen Olle Engkvist Byggm\"astare, and Ruth och Nils-Erik Stenb\"acks stiftelse.
DLL thanks Kyle Kaplan and Jacob McLane for help observing with IGRINS.
C. A. acknowledges partial support from the Spanish grant AYA2015-63588-P within the European Funds for Regional Development (FEDER). 
The Immersion Grating INfrared Spectrometer (IGRINS) was developed under a collaboration between the University of Texas at Austin and the Korea Astronomy and Space Science Institute (KASI) with the financial support from the W.J. McDonald Observatory, the US National Science Foundation under grant AST-1229522 to the University of Texas at Austin, and of the Korean GMT Project of KASI.

{\it Facilities: {Gemini Observatory}, {McDonald Observatory}, {ESO}}. 
\software{IRAF (Tody 1986, Tody 1993), Turbospectrum (Alvarez \& Plez 1998; Plez 2012), MOOG (Sneden 2012).}

\clearpage

\begin{table}
\centering
\caption{Observations, Adopted Stellar Parameters and Abundances}
\begin{tabular}{lcrccccccc}
\hhline{==========}
             &         &                & T$_{eff}$ & $\log$g & $\xi$               &            &          &          &     \\
             & Star    & Spectrograph   & (K)       & [cgs]   & (km$\cdot$s$^{-1}$) & [Fe/H]        & A(O)     & CN  \\ \hline
M4           & L1411   & Phoenix        & 3950  & 0.60 & 1.65            & -1.16$^a$  & 7.93$^a$ & Strong \\
             & L1501   & IGRINS         & 4150  & 0.85 & 1.50            & -1.15$^a$  & 7.83$^a$ & Strong \\
             & L1514   & Phoenix/IGRINS & 3875  & 0.35 & 1.95            & -1.18$^a$  & 8.18$^a$ & Weak   \\
             & L2307   & Phoenix        & 4075  & 0.85 & 1.45            & -1.15$^a$  & 7.91$^a$ & Strong \\
             & L2406   & IGRINS         & 4100  & 0.45 & 2.45            & -1.17$^a$  & 7.91$^a$ & Strong \\
             & L2410   & Phoenix        & 4550  & 2.45 & 1.50            & -1.18$^a$  & 8.10$^a$  & Weak \\
             & L3209   & Phoenix/IGRINS & 3975  & 0.60 & 1.75            & -1.17$^a$  & 8.00$^a$ & Weak   \\
             & L3413   & Phoenix        & 4175  & 1.20 & 1.65            & -1.13$^a$  & 8.16$^a$ & Weak   \\
             & L3624   & IGRINS         & 4225  & 1.10 & 1.45            & -1.11$^a$  & 8.06$^a$ & Weak   \\
             & L4401   & CRIRES         & 4700  & 2.80 & 1.40            & -1.18$^c$  &   ...    &       &       \\
             & L4404   & CRIRES         & 4650  & 2.10 & 1.50            & -1.18$^c$  & 7.89$^g$ & Strong   \\
             & L4511   & IGRINS         & 4150  & 1.10 & 1.55            & -1.11$^a$  & 7.98$^a$ & Strong \\
             & L4611   & Phoenix        & 3725  & 0.30 & 1.70            & -1.11$^a$  & 7.81$^a$ & Strong \\
             & L4613   & Phoenix        & 3750  & 0.20 & 1.65            & -1.14$^a$  & 7.81$^a$ & Weak   \\
             & L4630   & CRIRES         & 4145  & 1.30 & 2.00            & -1.18$^c$  &  ...  &     &    \\
             &         &                &           &          &                     &            &          &          &          \\
$\omega$ Cen & ROA 43  & Phoenix        & 3950  & 0.40 & 1.85                & -1.42$^f$  & 7.72$^f$ &   \\
             & ROA 132 & Phoenix        & 3900  & 0.30 & 1.85                & -1.32$^f$  & 7.91$^f$ &  \\
             & ROA 179 & Phoenix        & 3850  & 0.50 & 1.83                & -1.05$^f$  & 8.55$^f$ &  \\
             & ROA 213 & Phoenix        & 4500  & 1.00 & 1.73                & -2.12$^e$  & 7.35$^e$ &  \\
             & ROA 219 & Phoenix        & 3900  & 0.70 & 1.80                & -1.20$^e$  & 8.20$^e$ &  \\
             & ROA 324 & Phoenix        & 4000  & 0.70 & 1.75                & -0.90$^e$  & 8.28$^e$ &  \\ \hline
\end{tabular}
\begin{tablenotes}
\item \textbf{Notes}: (a) Ivans et al. (1999) (b) Smith et al. (2005) (c) de Laverny \& Recio-Blanco (2013) (d) Suntzeff \& Smith (1991) (e) Norris \& da Costa (1995) (f) Smith et al. (2000) (g) MacLean et al. (2018).
\end{tablenotes}
\label{tab:observations}
\end{table}

\begin{table}
\centering
\caption{Line list}
\begin{tabular}{ccccrcc}
\hhline{=======}
                   & $\lambda$ (Air)&      & $\chi$ &           & D$_o$  &            \\ 
Species / Molecule & (\AA)     & Line ID  & (eV)   & $\log$ gf & (eV)   & References \\ \hline
H$^{19}$F                 & 23358.329 & (1--0)R9  & 0.227  & -3.962    & 5.869  & J14;D00;ST84  \\
                   &           &           &        &           &        &      \\
Na I               & 23378.945 &           & 3.753  & -0.417    &        & R10  \\
                   & 23379.139 &           & 3.753  & 0.538     &        & R10  \\
                   &           &           &        &           &        &      \\
$^{12}$C$^{16}$O   & 23303.581 & (3--1)R70 & 1.421  & -4.517    & 11.092 & G94  \\
                   & 23304.591 & (3--1)R30 & 0.485  & -4.994    & 11.092 & G94 \\
                   & 23341.225 & (3--1)R26 & 0.431  & -5.065    & 11.092 & G94 \\
                   & 23367.752 & (2--0)R4  & 0.005  & -6.338    & 11.092 & G94 \\
                   & 23373.021 & (3--1)R23 & 0.396  & -5.124    & 11.092 & G94 \\
                   & 23374.033 & (3--1)R77 & 1.657  & -4.455    & 11.092 & G94 \\ \hline
\end{tabular}
\begin{tablenotes}
\item \textbf{Notes}: ST84 = Sauval \& Tatum (1984); D00 = Decin (2000); R10 = Ralchenko et al. (2010; NIST); G94 = Goorvitch (1994); J14 = J\"onsson et al. (2014a).
\end{tablenotes}
\label{tab:lines}
\end{table}

\clearpage

\begin{table}
\scalefont{0.7}
\centering
\caption{Abundance Results}
\begin{tabular}{lcrcccccccrr}
\hhline{============}
             & Star    & A(C)$_{R70}$ & A(C)$_{R30}$ & A(C)$_{R26}$ & A(C)$_{R4}$ & A(C)$_{R23}$ & A(C)$_{R77}$ & A(C)            & A(Na) & A(F)     & A(F)$_{corr}$ \\ \hline
             & Arcturus & 7.92        & 8.12         & 8.18         & 8.15        & 8.12         & 7.95         & 8.07 $\pm$ 0.10 & 5.97  & 3.63     & \\
M4           & L1411   &    ...       & 6.90         & 6.90         & 6.75        & 6.80         & 6.80         & 6.83 $\pm$ 0.06 & 5.45  & 2.50     & 2.58 \\
             & L1501   &    ...       & 7.15         & 6.95         & 7.00        & 6.85         & 7.00         & 6.99 $\pm$ 0.10 & 5.50  & $<$ 2.60 &  \\
             & L1514   &    ...       & 7.10         & 7.28         & 7.10        &   ...        &    ...       & 7.16 $\pm$ 0.08 & 5.03  & 2.67     & 2.82 \\
             & L2307   &   ...        & 7.12         & 7.12         & 6.90        & 7.12         &     ...      & 7.07 $\pm$ 0.10 & 5.40  & 2.60     & 2.55 \\
             & L2406   &  ...         & 6.70         & 6.85         & 6.87        & 6.78         &   ...        & 6.80 $\pm$ 0.07 & 5.23  & 2.65     & 2.57 \\
             & L2410   &   ...        &    ...       & 7.55         & 7.50        & 7.37         & 7.37         & 7.45 $\pm$ 0.08 & 5.15  & $<$ 3.80 & \\
             & L3209   &   ...        & 7.40         & 7.40         & 7.28        & 7.30         & 7.25         & 7.33 $\pm$ 0.06 & 5.24  & 2.62     & 2.66 \\
             & L3413   & 7.20         & 7.42         & 7.40         & 7.27        & 7.32         & 7.20         & 7.30 $\pm$ 0.09 & 5.03  & 2.85     & 2.69 \\
             & L3624   &   ...        & 7.60         & 7.50         & 7.35        & 7.43         & 7.60         & 7.50 $\pm$ 0.10 & 5.20  & 3.20     & 2.99 \\
             & L4401   & 7.40         & 7.50         & 7.45         & ...         & 7.35         & 7.55         & 7.45 $\pm$ 0.07 & 5.30  & $<$ 3.40 & \\
             & L4404   &  ...         & 7.07         & 7.27         & 7.07        & 7.20         &    ...       & 7.13 $\pm$ 0.08 & 5.60  & $<$ 3.50 & \\
             & L4511   &   ...        & 7.20         & 7.20         & 7.00        & 7.20         &     ...      & 7.15 $\pm$ 0.09 & 5.46  & 2.65     & 2.52 \\
             & L4611   & 6.95         & 7.07         & 7.15         & 7.05        & 7.20         & 6.92         & 7.06 $\pm$ 0.10 & 5.40  & 2.30     & 2.62 \\
             & L4613   &   ...        & 7.15         & 7.30         & 7.10        & 7.25         &   ...        & 7.20 $\pm$ 0.08 & 5.38  & 2.46     & 2.75 \\
             & L4630   & 7.10         & 7.02         & 7.05         & 6.95        & 7.02         & 7.15         & 7.05 $\pm$ 0.06 & 5.05  & 2.90     & 2.78 \\
             &         &              &              &              &             &              &              &                 &       &          & \\
$\omega$ Cen & ROA 43  &    ...       & 6.75         & 6.75         & 6.70        & 6.75         &    ...       & 6.74 $\pm$ 0.02 & 4.73  & $<$ 2.12 & \\
             & ROA 132 & 6.85         & 6.82         & 6.85         & 6.93        & 6.89         & 6.82         & 6.86 $\pm$ 0.04 & 4.62  & $<$ 2.46 & \\
             & ROA 179 & 7.82         & 8.15         & 8.23         & 8.08        & 8.15         & 7.72         & 8.03 $\pm$ 0.19 & 4.79  & $<$ 2.73 & \\
             & ROA 213 &   ...        &   ...       &  ...          & 7.10        & 7.05         &   ...        & 7.08 $\pm$ 0.03 & 4.21  & $<$ 2.66 & \\
             & ROA 219 &  ...         &    ...      &   ...         & 7.55        & 7.60         & 7.50         & 7.55 $\pm$ 0.04 & 4.94  & 2.80     & \\
             & ROA 324 &  ...         &   ...       &   ...         & 7.25        & 7.35         & 7.25         & 7.28 $\pm$ 0.05 & 6.25  & $<$ 2.85 & \\ \hline
\end{tabular}
\label{tab:abundances}
\end{table}

\clearpage

\begin{table}
\centering
\caption{Abundance Sensitivities to Stellar Parameters}
\begin{tabular}{cccccr}
\hhline{======}
element & $\delta$ T$_{eff}$ = +100 K & $\delta \log$g = +0.25 & $\delta$[Fe/H] = +0.1 & $\delta \xi$ = +0.3 km$\cdot$s$^{-1}$ & $\Delta$   \\ \hline
F       & +0.18        & +0.00    & +0.08        & +0.00                                & $\pm$ 0.20  \\
Na      & +0.10         &  -0.01     &  +0.01          & -0.02                               & $\pm$ 0.10  \\

C        &  +0.05   & +0.02   & +0.06      & -0.11       & $\pm$ 0.14   \\ \hline
\end{tabular}
\begin{tablenotes}
\item \textbf{Notes}: Baseline model:  T$_{eff}$ = 3725 K; log g = 0.30; [Fe/H] = -1.11; $\xi$ = 1.70 km/s. \end{tablenotes}
\label{tab:disturbance}
\end{table}

\clearpage

\begin{figure}[!htb]
   \centering
   \includegraphics[width=1.0\textwidth]{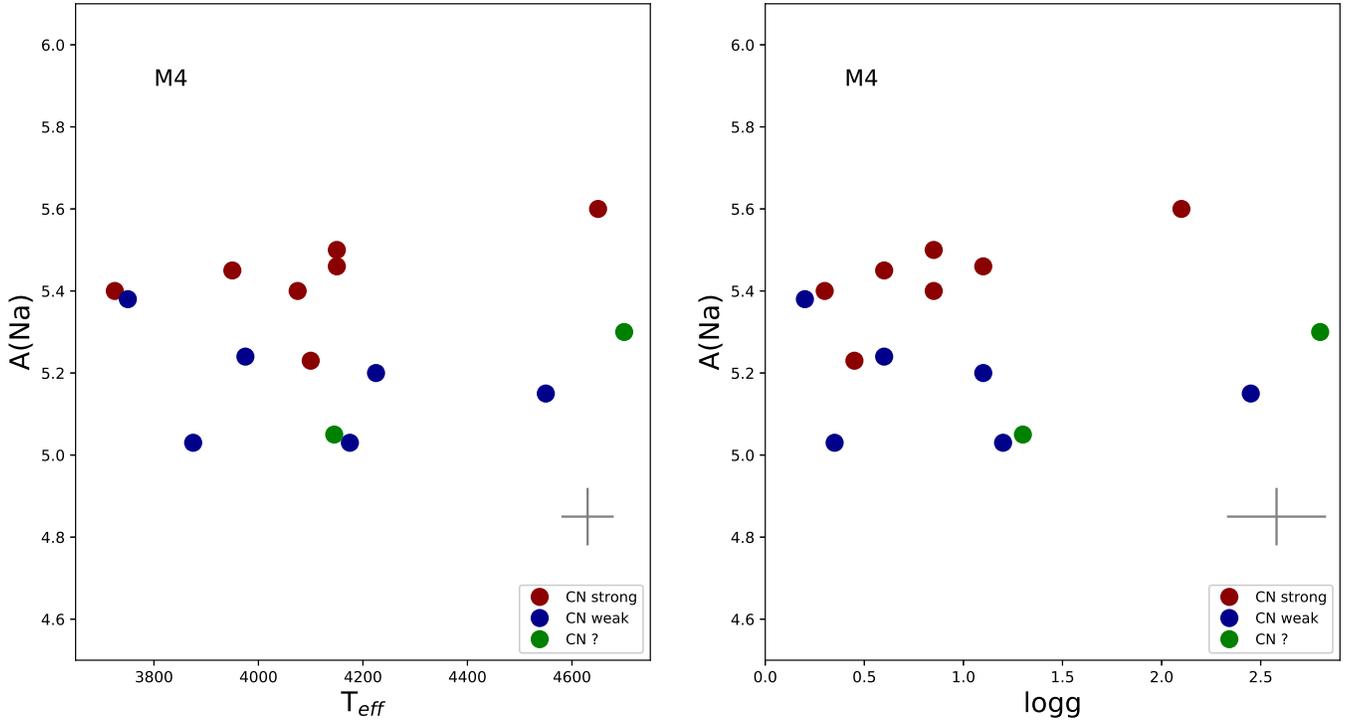}
   \caption{Sodium abundances for the M4 red giants analyzed in this work are presented versus T$_{eff}$ (left panel) and log g (right panel). The filled blue circles represent the CN-weak stars and the filled red circles represent the CN-strong stars, as classified in Ivans et al. (1999).
   Typical error bars are indicated in the figures.}
   \label{fig:Teff_logg_Na_M4}
\end{figure}

\begin{figure}[!htb]
   \centering
   \includegraphics[width=1.0\textwidth]{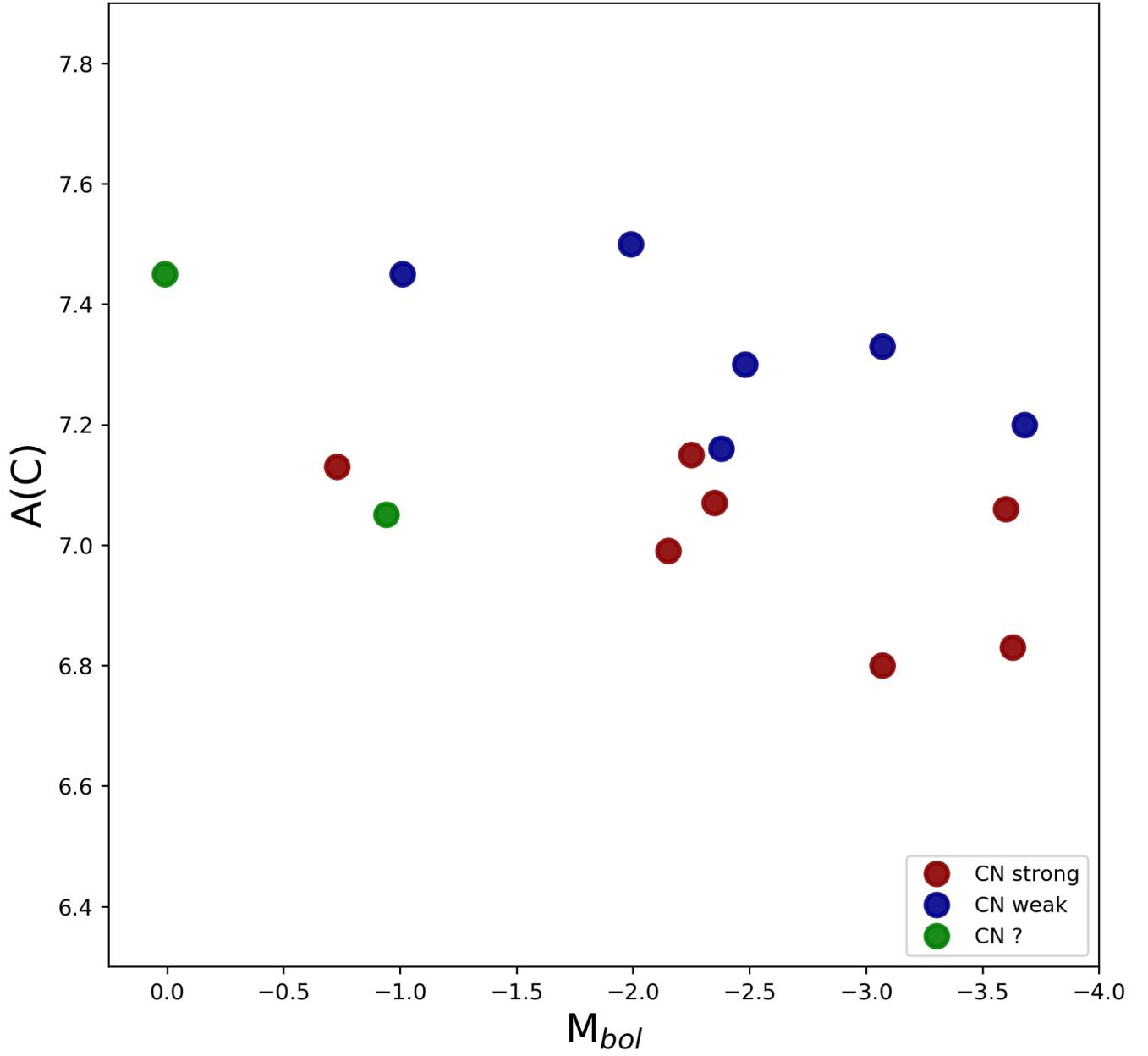}
   \caption{Derived carbon abundances versus the bolometric magnitudes (M$_{bol}$) for the studied M 4 red giants. 
   There is an overall decrease in the carbon abundances with increasing luminosity.
   The filled blue circles represent CN-weak stars and the filled red circles represent CN-strong stars, as classified in Ivans et al. (1999).
   }
   \label{fig:carbon_Mbol}
\end{figure}

\begin{figure}[!htb]
   \centering
   \includegraphics[width=1.0\textwidth]{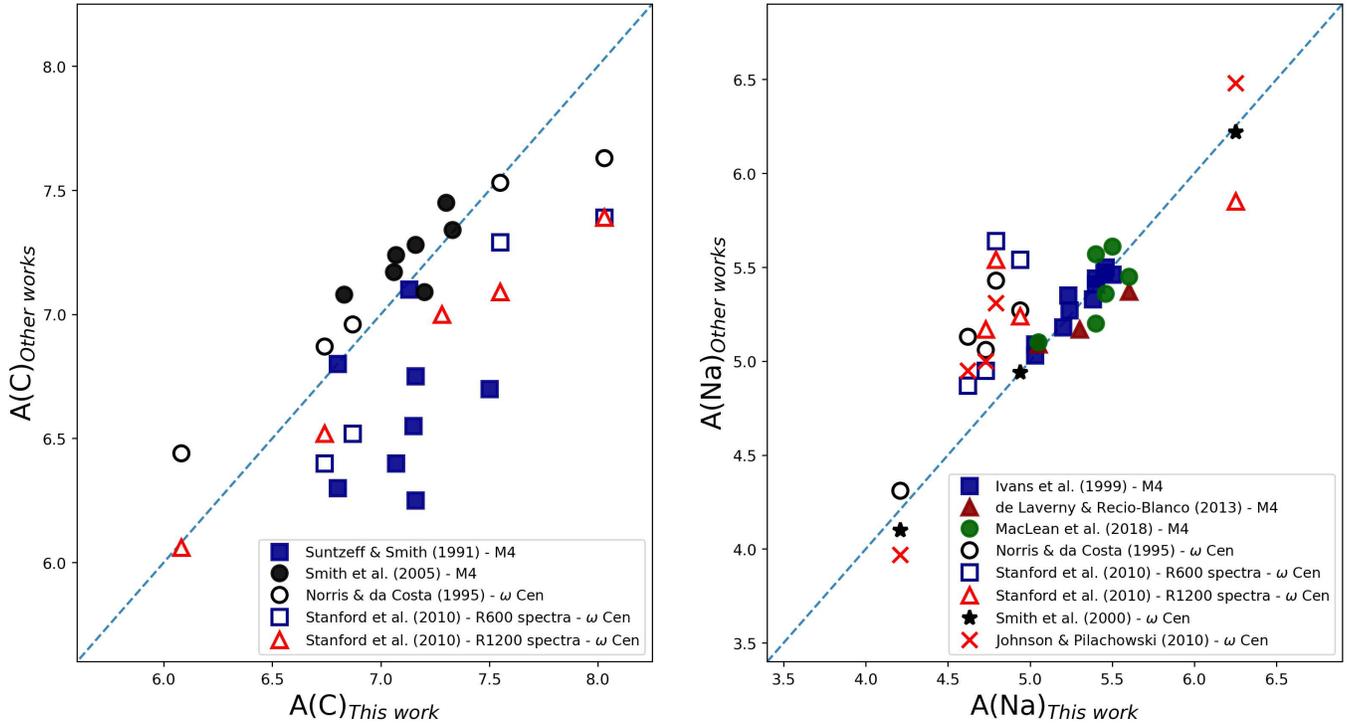}
   \caption{A comparison of the derived carbon (left Panel) and sodium (right Panel) abundances for M4 and $\omega$ Centauri stars with results from other studies in the literature obtained both from the optical and the near-infrared. $\omega$ Centauri stars are represented by open symbols, and M 4 stars by filled symbols.
  }
   \label{fig:carbon}
\end{figure}

\begin{figure}[!htb]
   \centering
   \includegraphics[width=1.0\textwidth]{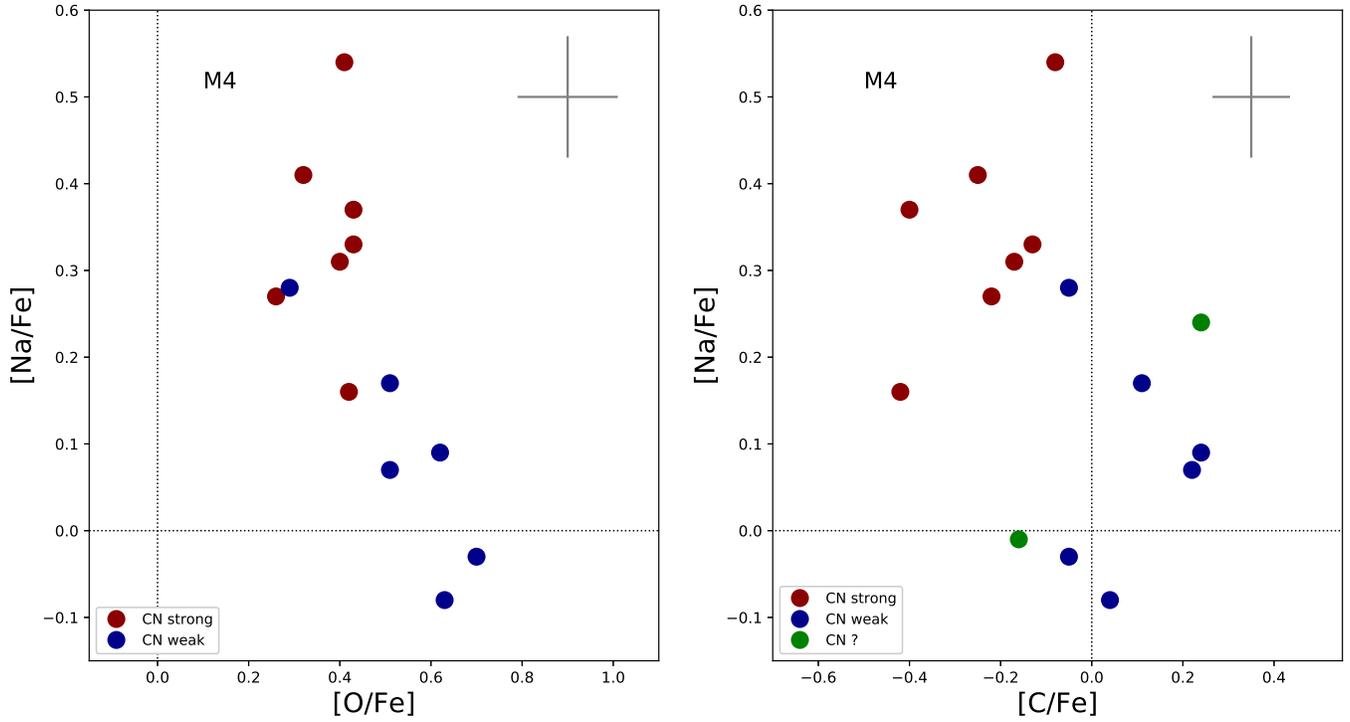}
   \caption{[Na/Fe] versus [O/Fe] (left panel) and [Na/Fe] versus [C/Fe] (right panel) for the studied red giants in M 4. The abundances show variations roughly within $\sim$0.5 dex, exhibiting a pattern of anti-correlation between Na-O and Na-C. 
   The oxygen abundances were taken from Ivans et al. (1999).
   Typical error bars are indicated in the figures.
   }
   \label{fig:C_O_Na_M4}
\end{figure}

\begin{figure}[!htb]
   \centering
   \includegraphics[width=1.0\textwidth]{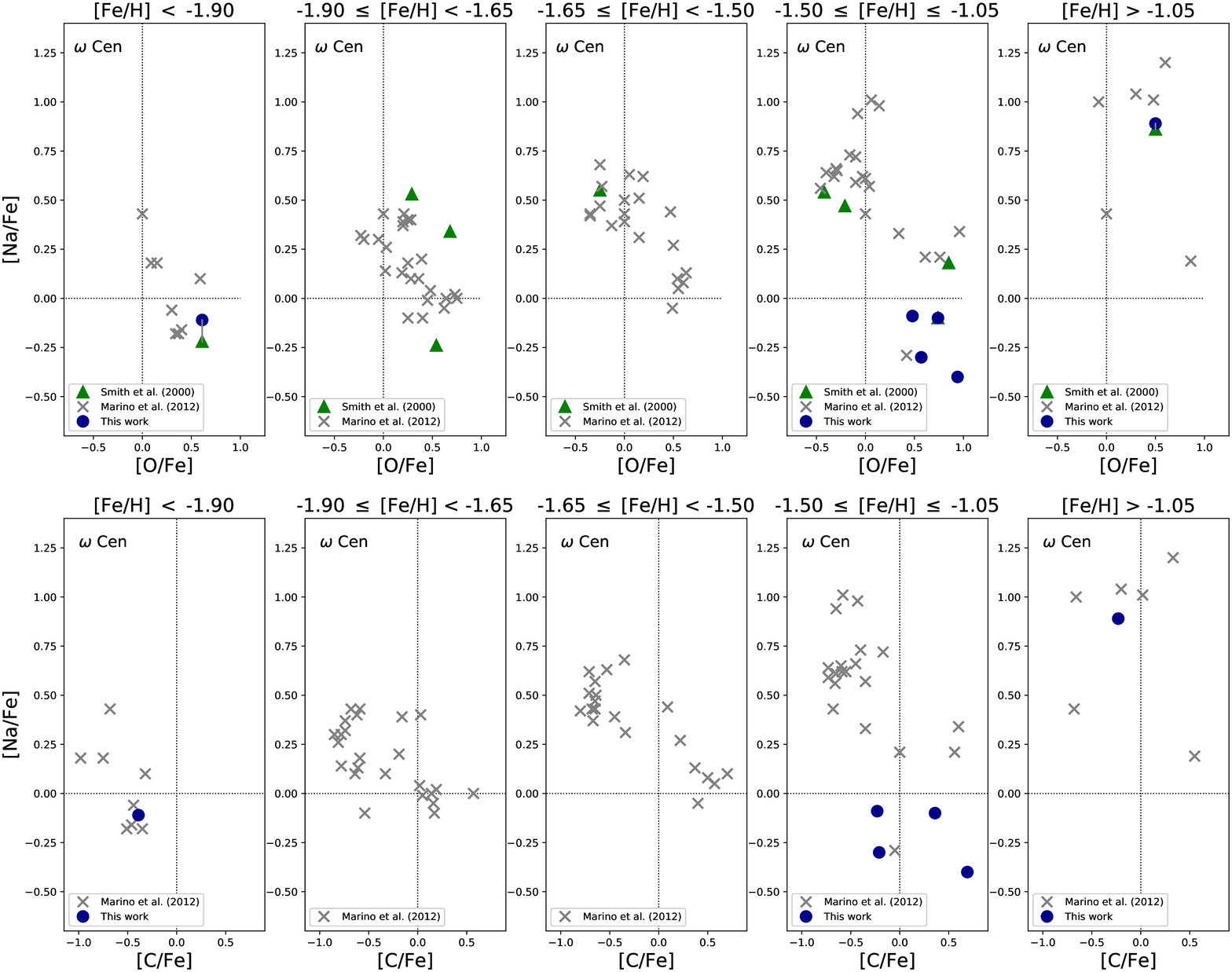}
   \caption{[Na/Fe] versus [O/Fe] (upper panels) and [Na/Fe] versus [C/Fe] (lower panels) abundance patterns for $\omega$ Centauri stars; the studied targets are segregated, according to their metallicities, into the five metallicity bins representing the $\omega$ Centauri populations in Marino et al. (2012): [Fe/H] $<$ -1.90, -1.90 $\leq$ [Fe/H] $<$ -1.65, -1.65 $\leq$ [Fe/H] $<$ -1.50, -1.50 $\leq$ [Fe/H] $\leq$ -1.05 and [Fe/H] $>$ -1.05. 
   The six targets studied in this work are shown as filled blue circles. We also show results from Smith et al. (2000).
   }
   \label{fig:O_vs_Na-C_vs_Na-o_Cen}
\end{figure}

\begin{figure}[!htb]
   \centering
   \includegraphics[width=1.0\textwidth]{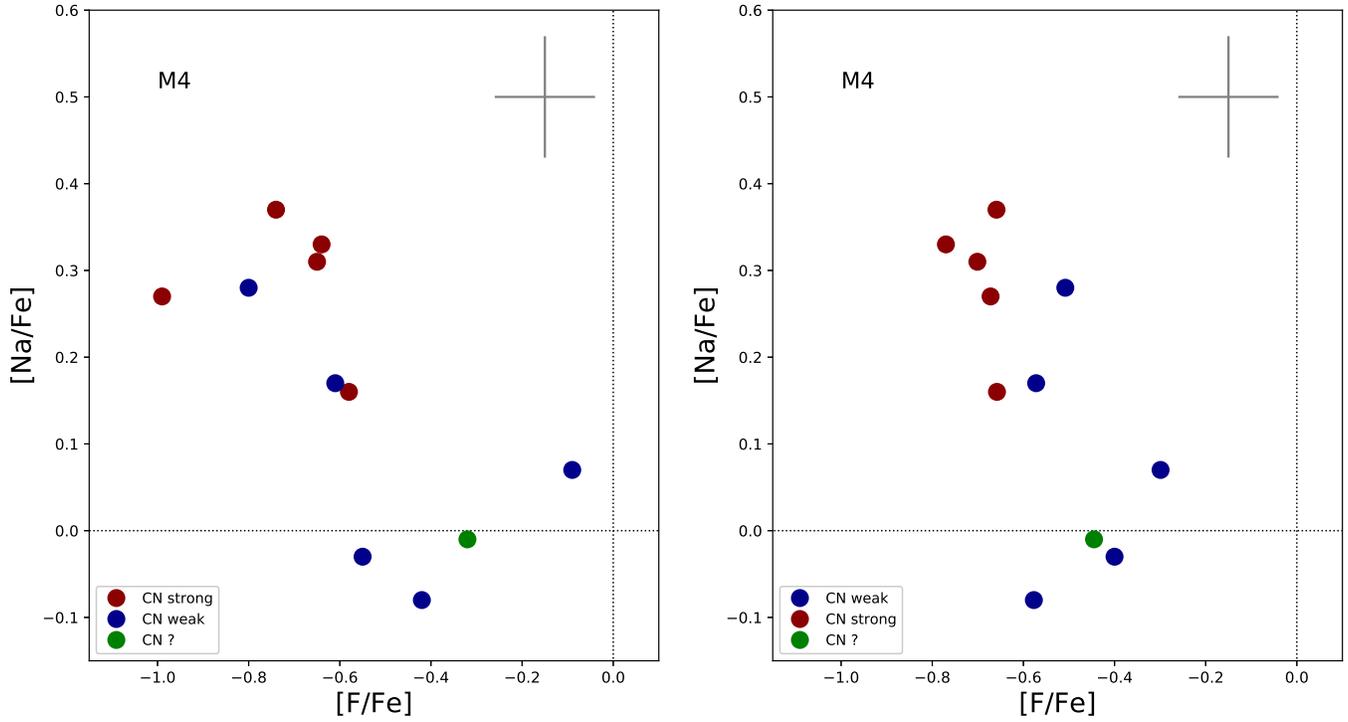}
   \caption{The anti-correlation between the [Na/Fe] and [F/Fe] abundances in M 4 obtained in this study. The target stars are represented by filled blue circles. Upper limits fluorine abundances have not been included.
   The derived fluorine abundances are shown in the left panel, while calibrated fluorine abundances corrected by empirically removing the observed trend between A(F) and stellar parameters (discussed in Section 5.1) are shown in the right panel. Typical error bars are indicated in the figures.
   }
   \label{fig:F_O_Na_M4}
\end{figure}

\begin{figure}[!htb]
   \centering
   \includegraphics[width=0.8\textwidth]{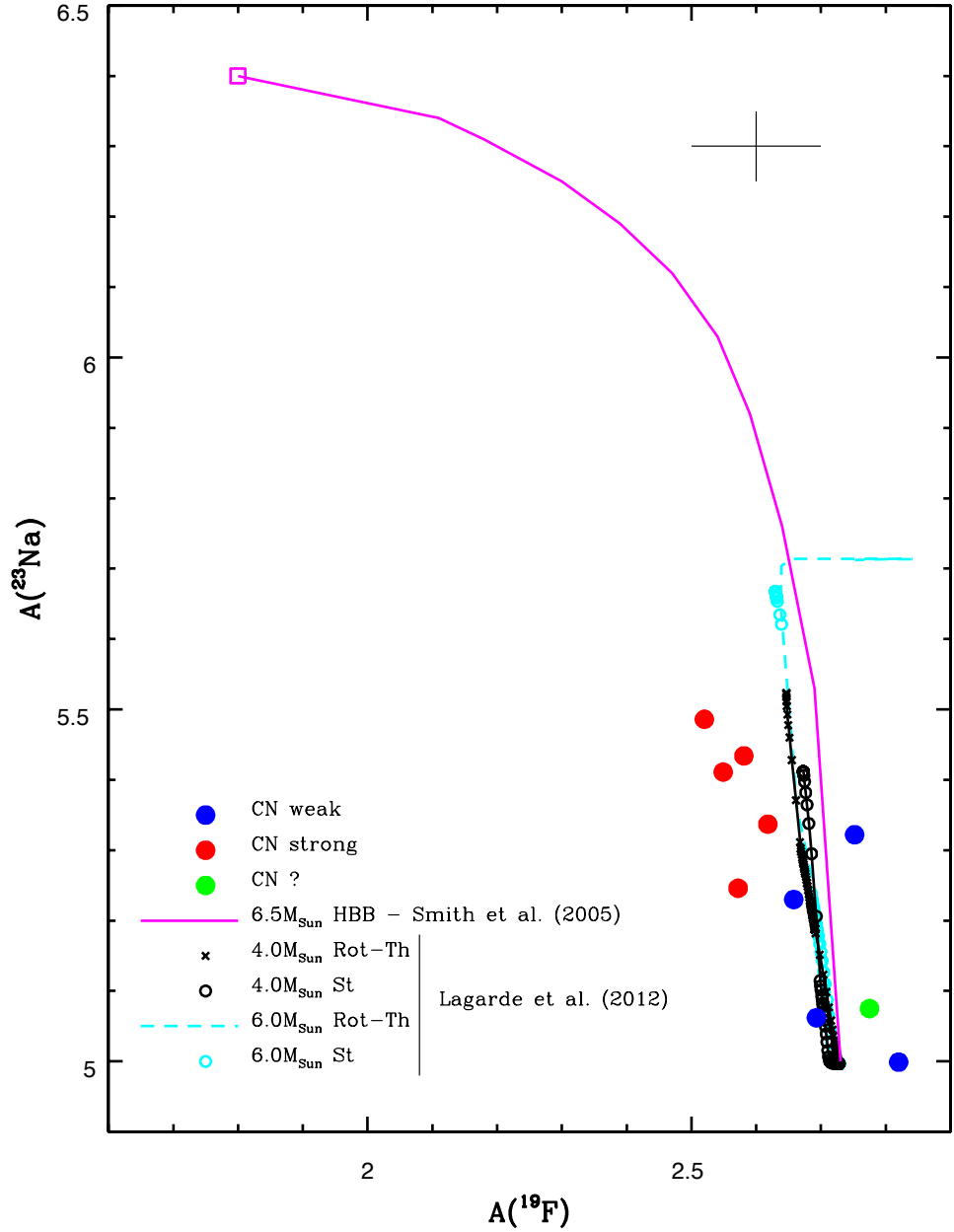}
   \caption{The sodium and corrected fluorine abundances (discussed in Section 5.1) in M 4 compared to abundances taken from stellar models presented in Smith et al. (2005) and Lagarde et al. (2012). 
   The open square at the end of the Smith et al. (2005) model track represents the final abundances of F and Na at the end of HBB on the AGB. The initial abundances for the models are taken to correspond to the F and Na abundances of the first generation of M 4 stars (the CN weak stars).  Although the models predict an anti-correlation between the abundances of F and Na, the slope of the A(Na)-A(F) anti-correlation is steeper in the observed abundances when compared to the model abundances. }
   \label{fig:F19vNa23All}
\end{figure}

\tablewidth{0pt}	

\end{document}